\documentclass[11pt]{article}

\usepackage{graphicx,amsfonts,amsbsy, amssymb, amsthm, amsmath}

\renewcommand{\vec}[1]{{\bf{#1}}}
\renewcommand{\Re}{{\mathfrak{Re}}}
\renewcommand{\Im}{{\mathfrak{Im}}}

\newcommand{\rmd}{{\mathrm d}}
\newcommand{\rme}{{\mathrm e}}
\newcommand{\rmi}{{\mathrm i}}
\newcommand{\Ord}{{\mathrm O}}
\newcommand{\littleo}{{\mathrm o}}

\newcommand{\boldpsi}{{\pmb{\psi}}}
\newcommand{\vecxi}{{\pmb{\xi}}}

\newcommand{\R}{{\mathbb R}}
\newcommand{\N}{{\mathbb N}}
\newcommand{\Q}{{\mathbb Q}}
\newcommand{\C}{{\mathbb C}}

\newcommand{\Prob}{{\mathbb P}}
\newcommand{\E}{{\mathbb E}}

\newcommand{\erf}{\mathop{\rm erf}}
\newcommand{\erfc}{\mathop{\rm erfc}}
\newcommand{\erfi}{\mathop{\rm erf{}i}}

\newcommand{\hash}{\#}
\newcommand{\eqcolon}{=:}
\newcommand{\coloneq}{:=}

\newcommand{\Lbar}{\bar{L}}
\renewcommand{\leq}{\leqslant}
\renewcommand{\geq}{\geqslant}
\renewcommand{\epsilon}{\varepsilon}
\renewcommand{\mod}[1]{\quad{\rm mod} \; #1}

\newtheorem{theorem}{Theorem}[section]
\newtheorem{proposition}[theorem]{Proposition}

\newtheorem{corollary}[theorem]{Corollary}
\newtheorem{lemma}[theorem]{Lemma}
\newtheorem{remark}[theorem]{Remark}

\newcommand{\dimostrazione}{\noindent{\sl Proof.}\phantom{X}}
\newcommand{\finire}{\hfill$\Box$}

\begin{document}

\title{No Quantum Ergodicity for Star Graphs}
\author{G.\ Berkolaiko${}^{1}$ \and
J.P.\ Keating${}^{2}$ \and B.\ Winn${}^{2}$\\
{\protect\small\em ${}^{1}$ Department of Mathematics,
University of Strathclyde, Glasgow G1 1XH, UK.}\\
{\protect\small\em ${}^{2}$School of
Mathematics, University of Bristol, Bristol BS8 1TW, UK.} }
\date{$25^{\rm th}$ July, 2003}

\maketitle
\begin{abstract}
We investigate statistical properties of the eigenfunctions of the 
Schr\"odinger operator on families of star graphs with incommensurate bond
lengths. We show that these 
eigenfunctions are not quantum ergodic in the limit as the number of bonds
tends to infinity by finding an observable for which  the 
quantum matrix elements do not converge to the classical average. We further
show that for a given fixed graph there are subsequences of eigenfunctions 
which
localise on pairs of bonds. We describe how to construct 
such subsequences explicitly. These constructions are analogous to scars
on short unstable periodic orbits.
\end{abstract}
\thispagestyle{empty}

\section{Introduction}\label{sec:intro}
Let $\psi_n$ denote the wave-function corresponding to the $n^{\rm th}$ 
energy level of a quantum system that has a Hamiltonian dynamical
system as its classical limit. We are interested in these wave-functions in 
the $n\to\infty$ limit, which corresponds to the semi-classical
regime. Numerical and some theoretical evidence supports the hypothesis that
their behaviour in this
limit is determined by general properties
of the underlying Hamiltonian such as, for example, time-reversibility, 
integrability and statistical properties of the flow 
(ergodicity, mixing, etc.).
A deeper understanding of this is one of the
goals of current research in quantum chaology.

When the classical Hamiltonian generates chaotic motion, 
the semi-classical eigenfunction hypothesis asserts that the wave-functions
should equidistribute over the appropriate energy shell \cite{B,V}. 
A physical explanation for this is that in the 
semi-classical
limit the quantum system should mimic the behaviour of the classical system;
if the classical motion is chaotic, then a typical trajectory ergodically 
explores the surface of constant energy in phase space. 
Another interpretation is that eigenstates 
are  invariant under time evolution, so it is natural to associate them
in the semi-classical limit with classical invariant sets.
One such invariant set is the energy shell itself.

The Schnirelman theorem \cite{Sc, CdV, Z, GL} states that 
for systems in which the Hamiltonian flow is ergodic the sequence of measures 
induced by $\psi_n$ converges to Liouville measure
in the limit as $n\to\infty$
{\em along a subsequence of density one}.
This behaviour has been termed 
``quantum ergodicity''. Quantum ergodicity implies a weak version
of the semi-classical eigenfunction hypothesis \cite{BSS}.

It is possible that quantum ergodic systems have subsequences
of states for which the corresponding measures do not converge to Liouville
measure (of course such subsequences have density zero). 
These subsequences, if they exist, are expected to 
 be associated with other classical invariant sets,
such as periodic orbits. The case where the limit of an exceptional 
subseqence is a singular measure supported on one-or-more isolated, unstable
periodic orbits of the classical system is called ``scarring''.

Scarred eigenstates were observed numerically by Heller \cite{H}, who proposed
the first theoretical explanation for their existence, based on the
semi-classical evolution of a wave-packet centred on a periodic orbit 
under linearised dynamics. 
Another important development was an understanding of
the contribution to wave-functions from {\em all} periodic orbits 
\cite{Bog, B2} resulting in
formul\ae{} related to the semiclassical trace formula for the density of
states. 
Later, the theory was extended to include non-linear effects
\cite{KH} and, more recently, 
situations where the orbit in question undergoes a bifurcation
\cite{KP}. A review of related works was given in \cite{Ka}.
All of the above mentioned theories relate to scar effects in averages over
a semiclassically increasing number of states. 
This may be thought of as a weakened
form of scarring, because it is not clear that any one state in the averaging
range causes the scar; the scars may be a collective effect.
It is a much harder problem to show that a particular sequence of individual
states is scarred. Currently, the only systems
known rigorously to support scarring in this strong form
are the cat maps \cite{FNdB} which have non-generic spectral
statistics caused by number-theoretical symmetries \cite{K}.

For quantum graphs \cite{KS1} the wave-functions are the eigenfunctions
of the (continuous) Laplace operator on the bonds 
with matching conditions at the vertices 
chosen to make the problem self-adjoint.
There is evidence to suggest that the spectral statistics of
large quantum graphs coincide with those of generic quantised, 
classically chaotic, systems \cite{KS2, BSW1, BSW2, Berk} subject to 
mild conditions on the connectivity \cite{T}.
Although the classical dynamics on a graph are not Hamiltonian,
they are ergodic and
it might be expected that an analogue of the Schnirelman theorem should
hold.

Graphs are a rich source of problems in quantum chaology
and related fields. Recent works have considered:
scattering problems \cite{KS3, TM, KS4},
the spacing distribution of eigenvalues \cite{BG},
nodal domain statistics \cite{GSW},
the  Dirac operator on graphs \cite{BoHa1,BoHa2},
Brownian motion on graphs \cite{CDM, D}
and the important question of how to construct
families of graphs with increasing numbers of bonds \cite{PTZ}.

Recently, authors have begun to investigate the wave-functions
of quantum graphs. 
Kaplan \cite{K2} studied eigenfunction statistics for ring-graphs 
using a combination of numerical techniques and analytical calculations
of the short-time semiclassical behaviour of a wave-packet 
close to a 1-bond periodic orbit. 
The inverse participation ratio (a measure of localisation in a given
state)  was found to be well-described by this contribution, and shows
deviation from the ergodically expected behaviour.
Similar deviations were noticed for lattice-graphs. Remarkably,
Schanz and Kottos \cite{SK} observed that it would be impossible for
the shortest orbits that are responsible for this enhanced localisation
to support strong scarring. They wrote down an
explicit criterion
which must be satisfied by the energy of any strongly scarred state,
and deduced asymptotics of the probability distribution
of scarring strengths.
In \cite{KMW} a study was made of the eigenfunctions of a family of graphs
known as star graphs (the
name being derived from the  connectivity of graphs in the family).
The value distribution for the amplitude of eigenfunctions on a 
single bond of the graph, subject to an appropriate normalisation, was
rigorously calculated in the limit as the number of bonds tends to 
infinity. In fact the normalisation implies that star graphs with a 
fixed, {\em finite} number of bonds are not
quantum ergodic. 
However, this result leaves open the question of whether
star graphs are quantum ergodic in the limit as the number of bonds
tends to infinity. This is because one bond represents a vanishingly small 
fraction
of a graph when the number of bonds becomes infinite,
whereas quantum ergodicity is concerned with structures on 
macroscopic (classical) scales.

The results we present here extend the work in \cite{KMW} on star graphs.
We review the definition
of a quantum star graph in section \ref{sec:2} below.
We show that (see the following subsection for precise statements) 
quantum star graphs are not quantum ergodic in the limit as the
number of bonds tends to infinity. We also show that for any
given star graph there exist exceptional
subsequences of eigenfunctions that become localised on pairs of 
bonds as $n\to\infty$. Orbits on a graph are simply itineraries of bonds,
so this localisation is analogous to strong scarring on short period-2 orbits.
Such orbits are unstable in the sense that there is an exponentially small 
probability of remaining on a given orbit.
Our explicit construction supports the observation
of Schanz and Kottos \cite{SK} that star graphs support a 
large number of states scarred in such a way.

The spectral statistics of
star graphs are different to those associated with the more general graphs
described above \cite{BK}. 
The fact that quantum star graphs are not quantum 
ergodic does not contradict the possibility of a quantum ergodicity
theorem for graphs with general connectivity.
It is known that the spectral statistics of quantum star graphs are
the same as those associated with
the family of \v{S}eba billiards \cite{Se,BBK}, so-called ``intermediate
statistics''. There is evidence to suggest
that the results we present on scarring can also be extended to \v{S}eba
billiards \cite{BKW}.
\subsection{Main results}

To investigate quantum ergodicity for large star graphs, we consider 
an observable that picks out a positive proportion of the graph.
We consider a graph with $\alpha v$ bonds, where $\alpha, v\in\N$, and
the observable $\vec{B}=(B_i(x))_{i=1}^{\alpha v}$ defined by
\begin{equation}
  B_i\coloneq\left\{
  \begin{array}{ll}
    1 & \mbox{for $i=1,\ldots,v$} \\
    0 & \mbox{for $i=v+1,\ldots,\alpha v$.}
  \end{array} \right.
\label{eq:def:B}
\end{equation}
$\vec{B}$ may be thought of as the indicator function of the first $v$ bonds.
The classical average of $\vec{B}$ is approximately $1/\alpha$.
We shall consider the limit $v\to\infty$.

Wave-functions on graphs have a component on each bond, so we shall use the 
notation
\begin{equation*}
  \boldpsi^{(n)}\coloneq(\psi_i^{(n)})_{i=1}^{\alpha v}
\end{equation*}
for the $n^{\rm th}$ eigenstate. The inner product 
$\langle\cdot|\cdot\rangle$ is defined in \eqref{eq:inprod} below.

Each bond of the graph has a length, and the vector of bond lengths
will be denoted $\vec{L}\coloneq(L_i)_{i=1}^{\alpha v}$.

\begin{figure}[h]
\begin{center}
\includegraphics[angle=0,width=8.0cm,height=6cm]{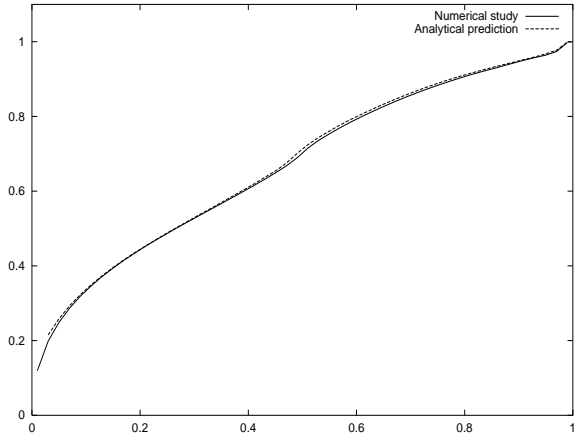}
\caption{Comparing numerical data with the analytical prediction, $F(R)$.  
For this plot $\alpha=3$ and in the numerical study, $v=30$.}
\label{fig:5}
\end{center}

\end{figure}

\begin{theorem} \label{thm:2}
  For each $v$
let the components of $\vec{L}$ be linearly independent over $\Q$. Then
there exists a probability density $p_v(\eta)$ such that for any 
continuous function $h$,
\begin{equation}
  \lim_{N\to\infty}\frac{1}{N}\sum_{n=1}^{N} 
h( \langle\boldpsi^{(n)}|\vec{B}|\boldpsi^{(n)}\rangle )=
\int_{-\infty}^{\infty}h(\eta)p_v(\eta)\rmd\eta.
\end{equation}
The density $p_v(\eta)$ is supported on the interval $[0,1]$.
\end{theorem}
\begin{theorem} \label{thm:3}
  For each $v$ let the bond lengths $L_j$, $j=1,\ldots,\alpha v$ lie in the 
range $[\Lbar,\Lbar+\Delta L]$ and be linearly independent
over ${\mathbb Q}$. If $v\Delta L\to0$ as $v\to\infty$
then there exists a probability distribution function $F(R)$ such
that for any $R\in(0,1)$,
\begin{equation}
  \lim_{v\to\infty}\int_{-\infty}^R p_v(\eta)\rmd\eta = F(R)
\end{equation}
where 
\begin{equation}
  F(R)=\frac{1}{2}-
\left.\frac{1}{\pi\alpha}\Re\int_{-\infty}^{\infty} P_\eta(\xi)
\left(\arg(\tau_\eta(\xi))-\rmi\log|\tau_{\eta}(\xi)|\right)\rmd\xi
\right|_{\eta=1/R-1},
\end{equation}
and
\begin{align*}
 P_\eta(\xi)=&\frac{1}{\sqrt{\pi\eta}}\exp\left(\frac{-\rmi\pi}{4}+
\frac{\rmi\xi^2}{4\eta}\right)+\frac{(\alpha-1)}{\sqrt{\pi}}\exp
\left(\frac{\rmi\pi}{4}-\frac{\rmi\xi^2}{4}\right),\\
  \tau_\eta(\xi)=& \frac{2}{\sqrt{\pi}}\sqrt{\eta}\exp\left(
\frac{\rmi\pi}{4}+\frac{\rmi\xi^2}{4\eta}\right)+\xi\erf\left(
\frac{\rme^{-\rmi\pi/4}\xi}{2\sqrt{\eta}}\right) \nonumber \\
&+\frac{2(\alpha-1)}{\sqrt{\pi}}\exp\left(
-\frac{\rmi\pi}{4}-\frac{\rmi\xi^2}{4}\right)+\xi(\alpha-1)\erf\left(
\frac{\rme^{\rmi\pi/4}\xi}{2}\right).
\end{align*}
\end{theorem}
\noindent
The function $F(R)$ is plotted in figure \ref{fig:5}.
\begin{remark} \label{rem:1:3}
  If star graphs satisfied quantum ergodicity, then $F(R)$ would be
the step-function
  \begin{equation}
    F(R)=\left\{ 
    \begin{array}{ll}
      1, & \mbox{for $R>1/\alpha$} \\
      0, & \mbox{for $R<1/\alpha$}
    \end{array}
\right.
  \end{equation}
for this observable.
\end{remark}

\begin{figure}[h]
\begin{center}
\includegraphics[angle=0,width=8.0cm,height=6cm]{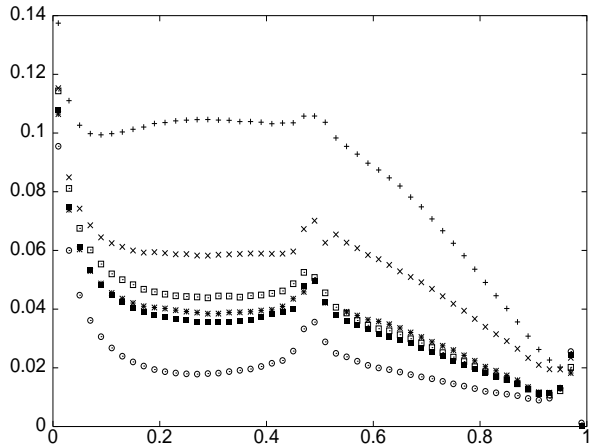}
\caption{Convergence to $F(R)$ for $v=5 (+), 10 (\times), 
15 (+\mspace{-14mu}\times), 
20 (\boxdot), 25 (\blacksquare), 30 (\odot)$.}
\label{fig:6}
\end{center}
\end{figure}

In figure \ref{fig:5} we compare the numerical data for the value distribution
 of 
$\langle\boldpsi^{(n)}|\vec{B}|\boldpsi^{(n)}\rangle$  for a star graph with
90 bonds with the $v\to\infty$ analytical prediction $F(R)$. 
The difference between the actual distribution $F(R)$ and that which would be 
expected if the graph were quantum ergodic (remark \ref{rem:1:3}) is clear.
Figure \ref{fig:6} shows the difference between numerical data and $F(R)$ for
increasing values of $v$.

We we also show that for graphs with fixed number of bonds, there are 
subsequences of eigenfunctions that localise on two bonds.
\begin{theorem} \label{thm:1}
  Let the elements of $\vec{L}$ be linearly independent over
  $\Q$. Given any distinct two bonds, indexed by $i_1$ and $i_2$, of a $v$-bond
  star graph, there exists a subsequence $(k_{n_r})\subseteq(k_n)$
  such that for any $\vec{f}=(f_i)_{i=1}^{v}$ smooth in each
  component,
  \begin{equation}
    \label{eq:scars_thm}
    \lim_{r\to\infty}  
    \langle \boldpsi^{(n_r)} | \vec{f} | \boldpsi^{(n_r)} \rangle
    = \frac{1}{L_{i_1}+L_{i_2}} \left( 
      \int_0^{L_{i_1}} f_{i_1}(x)\rmd x 
      + \int_0^{L_{i_2}} f_{i_2}(x)\rmd x
    \right).
  \end{equation}
\end{theorem}



\section{Quantum Star Graphs} \label{sec:2}
\begin{figure}[h]
\begin{center}
\includegraphics[angle=0,width=3.0cm,height=2.5cm]{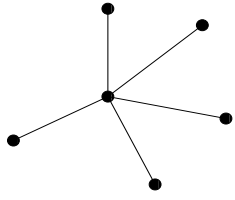}
\caption{A star graph with 5 bonds}
\label{fig:0}
\end{center}
\end{figure}

A star graph\footnote{sometimes referred to as a Hydra graph} is a metric 
graph with $b$ vertices all connected only to one central vertex. Thus there
are $b+1$ vertices and $b$ bonds (figure \ref{fig:0}). We shall denote by
$\vec{L}\in\R^b$ the vector of bond lengths.

We define the quantum star graph in the following way. Let $\mathcal{H}$ 
denote the real Hilbert space
\begin{equation}
  \mathcal{H}\coloneq L^2([0,L_1])\times\cdots\times L^2([0,L_b])
\end{equation}
with inner product
\begin{equation} \label{eq:inprod}
   \langle\vec{f}|\vec{g}\rangle\coloneq
  \sum_{j=1}^b\int_0^{L_j}f_j(x)g_j(x)\rmd{x}.
\end{equation}
Elements of $\mathcal{H}$ are denoted $\vec{f}=(f_1,\ldots,f_b)$.
Let $\mathcal{F}\subseteq\mathcal{H}$ be the subset of functions 
$\vec{f}$ which are twice-differentiable in
each component and satisfy the conditions
\begin{eqnarray}
f_j(0)=f_i(0)&\eqcolon& f_0,\qquad j,i=1,\ldots,b \label{eq:bc1}\\
\sum_{j=1}^b f_j^{\prime}(0)&=&\frac{1}{\lambda}f_0\\
f_j^{\prime}(L_j)&=&0,\qquad j=1,\ldots,b.  \label{eq:bc2}
\end{eqnarray}
The parameter $\lambda$ may be varied to give different boundary conditions
at the central vertex of the graph. Henceforth we shall concentrate on
the case $1/\lambda=0$, the so-called Neumann condition.
The Laplace operator $\triangle$ on $\mathcal{F}$ is defined by
\begin{equation}
  \triangle \vec{f}\coloneq\left(\frac{\rmd^2f_1}{\rmd x^2},\ldots,
\frac{\rmd^2{f_b}}{\rmd x^2}\right).
\end{equation}
$\triangle$ defined on $\mathcal{F}$ is self-adjoint. Since the
space on which the functions in $\mathcal{F}$ are defined is compact,
the operator $\triangle$ has a discrete spectrum of eigenvalues (\cite{DS}, 
Section XIII.4).
i.e.\ the equation
\begin{equation} \label{eq:schrodinger}
  -\triangle\boldpsi=k^2\boldpsi
\end{equation}
has non-trivial solutions for $k=k_1, k_2,\ldots$ Such $\boldpsi$ are the
wave-functions \cite{KS1, KS2}. We shall use the notation that 
$\boldpsi^{(n)}\coloneq(\psi_i^{(n)}(x))_{i=1}^{b}$ is the wave-function
corresponding to $k=k_n$.

Solving \eqref{eq:schrodinger} with boundary conditions 
\eqref{eq:bc1}--\eqref{eq:bc2}, we find that the component of the
$n^{\rm th}$ normalised eigenfunction of the Laplace operator on 
the $i^{\rm th}$ bond of a star graph is
\begin{equation}
\label{eq:efun}
  \psi_i^{(n)}(x)=A_i^{(n)} \cos k_n(x-L_i)
\end{equation}
where the amplitude is given by
\begin{equation}
\label{eq:norm}
  A_i^{(n)}=\left(
    \frac{2\sec^2k_nL_i}{\sum_{j=1}^{b} L_j\sec^2{k_nL_j}}
  \right)^{1/2}
\end{equation}
and $k_n$ is the $n^{\rm th}$ positive solution to
\begin{equation}
  Z(k,\vec{L})\coloneq\sum_{j=1}^{b}\tan k L_j = 0.
\end{equation}

In sections \ref{sec:3}--\ref{sec:6} it will be convenient to take
$b=\alpha v$, where $\alpha\in\N$ is fixed. This is so that we can
easily describe a fraction of the total number of bonds as the number of bonds
becomes large ($v\to\infty$).
In section \ref{sec:7} we shall take $\alpha=1$ for notational
convenience, since there we will only be concerned with fixed graphs.

\section{Distribution of the observable $\vec{B}$} \label{sec:3}

In this section we prove the existence of a limit distribution for
the diagonal matrix elements of $\vec{B}$ on star graphs with a fixed number, 
$\alpha v$ of bonds.

\begin{lemma} \label{lem:2:1}
Consider a star graph with $\alpha v$ bonds with fixed lengths given
by the vector $\vec{L}$.
Then for $\vec{B}$ defined by \eqref{eq:def:B},
  \begin{equation}
    \langle\boldpsi^{(n)}|\vec{B}|\boldpsi^{(n)}\rangle = 
\frac{\sum_{i=1}^v L_i\sec^2{k_nL_i}}{\sum_{j=1}^{\alpha v}L_j\sec^2 k_nL_j}
+\Ord\left(\frac{1}{k_n}\right)
  \end{equation}
where the error estimate is uniform in $v$ and $L_i\geq L_{\rm min}>0$ for
each $i$.
\end{lemma}
\dimostrazione 
We recall that
\begin{equation}
   \langle\boldpsi^{(n)}|\vec{B}|\boldpsi^{(n)}\rangle= \sum_{j=1}^{\alpha v}
\int_0^{L_j} |\psi_j^{(n)}(x)|^2B_j(x)\rmd x.
\end{equation}
Integrating \eqref{eq:efun} gives, for $i=1,\ldots,v$,
\begin{equation}
  \int_0^{L_i} |\psi_i^{(n)}(x)|^2 B_i(x)\rmd x =
\frac{1}{\sum_{j=1}^{\alpha v} L_j\sec^2{k_nL_j}}\left(
L_i\sec^2 k_nL_i +\frac{1}{k_n}\tan k_nL_i\right),
\end{equation}
and for $i\geq v+1$,
\begin{equation}
  \int_0^{L_i} |\psi_i^{(n)}(x)|^2 B_i(x)\rmd x=0.
\end{equation}
Thus
\begin{equation}
  \langle\boldpsi^{(n)}|\vec{B}|\boldpsi^{(n)}\rangle =
\frac{\sum_{i=1}^v L_i\sec^2{k_nL_i}}{\sum_{j=1}^{\alpha v}L_j\sec^2 k_nL_j}
 + \frac{E}{k_n}
\end{equation}
where
\begin{equation}
  E=\frac{\sum_{i=1}^v \tan{k_nL_i}}{\sum_{j=1}^{\alpha v}L_j\sec^2 k_nL_j}.
\end{equation}
Let $L_{\rm min}\coloneq\min_j\{L_j\}$. Then
\begin{eqnarray}
  |E|&\leq&\frac{\sum_{i=1}^v |\tan{k_nL_i}|}
{(\alpha-1)L_{\rm min}v+L_{\rm min}\sum_{j=1}^{v}\sec^2 k_nL_j}\\
&\leq&\left( L_{\rm min}+\frac{(\alpha-1)vL_{\rm min}}{\sum_{i=1}^{v}
\sec^2{k_n L_j}}\right)^{-1}
\end{eqnarray}
using the fact that $|\tan\theta|\leq\sec^2\theta$ for any $\theta\in\R$.
Hence $E=\Ord(1)$ as $n\to\infty$ uniformly in $v$, $L_{\rm min}>0$. \finire


\noindent{\sl Proof of theorem \ref{thm:2}.}\phantom{X}
By lemma \ref{lem:2:1},
\begin{equation}
 h( \langle\boldpsi^{(n)}|\vec{B}|\boldpsi^{(n)}\rangle)
 =h\left(\frac{\sum_{i=1}^v L_i\sec^2{x_i}}
{\sum_{j=1}^{\alpha v}L_j\sec^2 x_j}\right)+E_n
\end{equation}
where $E_n=\littleo(1)$ as $n\to\infty$ since $h$ is
uniformly continuous on $[0,1]$. 
Hence
\begin{equation}
  \frac{1}{N}\sum_{n=1}^{N} E_n\to0\quad\mbox{as $N\to\infty$.}
\end{equation}
Therefore
\begin{equation}
  \lim_{N\to\infty}\frac{1}{N}\sum_{n=1}^N
h( \langle\boldpsi^{(n)}|\vec{B}|\boldpsi^{(n)}\rangle)=
\lim_{N\to\infty}\frac{1}{N}\sum_{n=1}^N
h\left(\frac{\sum_{i=1}^v L_i\sec^2{k_nL_i}}
{\sum_{j=1}^{\alpha v}L_j\sec^2 k_nL_j} \right).
\end{equation}
According to Barra and Gaspard, there is an absolutely continuous 
measure $\nu(\vecxi)$ such
that for piecewise continuous functions, $f:\Sigma\to\R$,
\begin{equation*}
  \lim_{N\to\infty}\frac{1}{N}\sum_{n=1}^{N} f(k_n\vec{L})=
\int_{\Sigma} f(\vecxi)\rmd\nu(\vecxi)
\end{equation*}
where $\Sigma$ is the surface embedded in the $\alpha v$ dimensional torus 
with side $\pi$,  defined by
\begin{equation*}
  \tan x_1+\cdots+\tan x_{\alpha v}=0.
\end{equation*}
$\vecxi$ is a set of $\alpha v-1$ coordinates which parameterise $\Sigma$.
To avoid repetition,
we refer the reader to \cite{BG}, \cite{KMW} for more detail about this result
and its application to similar problems.

Let
\begin{equation}
  f(\vec{x})=h\left(\frac{\sum_{i=1}^v L_i\sec^2{x_i}}
{\sum_{j=1}^{\alpha v}L_j\sec^2 x_j}\right)
\end{equation}
we can define $p_v(\eta)$ by
\begin{equation}
  \int_{\Sigma} f(\vecxi)\rmd\nu(\vecxi)\eqcolon 
\int_{-\infty}^{\infty}
h(\eta)p_v(\eta)\rmd\eta.
\end{equation}

Since $0\leq \langle\boldpsi^{(n)}|\vec{B}|\boldpsi^{(n)}\rangle \leq1$,
it follows that $p_v(\eta)$ is supported on $[0,1]$.\finire

\section{The large graph limit}
Let $\eta\in\R$ and define
\begin{equation}
  X_\eta(n)\coloneq\frac{1}{v^2}\sum_{j=v+1}^{\alpha v}L_j\sec^2 k_nL_j-
\frac{\eta}{v^2}\sum_{i=1}^v L_i\sec^2k_nL_i
\end{equation}
for $n=1,2,\ldots\;$ The key result of this section is the following.
\begin{proposition} \label{prop:5:2}
For each $v$, there exists a probability density function $f_{X_\eta,v}$
such that
\begin{equation}
  \lim_{N\to\infty}\frac{1}{N}\hash\left\{ n\in\{1,\ldots,N\}: X_{\eta}(n)<S
\right\}=\int_{-\infty}^{S} f_{X_\eta,v}(\sigma)\rmd\sigma.
\end{equation}
Furthermore, for each $S\in\R$,
\begin{equation*}
  \int_{-\infty}^{S} f_{X_\eta,v}(\sigma)\rmd\sigma\to
\int_{-\infty}^{S} f_{X_\eta}(\sigma)\rmd\sigma
\end{equation*}
as $v\to\infty$, provided that $v\Delta L\to0$ in this limit, where
\begin{equation*}
  f_{X_\eta}(\sigma)=\frac{-1}{4\alpha\sqrt{\pi}}\Re\int_{-\infty}^{\infty}
P_\eta(\xi)\frac{
\rme^{3\rmi\pi/4}\tau_\eta(\xi)}{(-\sigma)^{3/2}}
w\left(\frac{\rme^{3\rmi\pi/4}\tau_\eta(\xi)}{2\sqrt{-\sigma}}
\right)\rmd\xi.
\end{equation*}
The functions $P_\eta$ and $\tau_\eta$ are defined by
\eqref{eq:def:P} and \eqref{eq:def:tau} below, and
$w(z)\coloneq\rme^{-z^2}\erfc(-\rmi z)$.
\end{proposition}
\dimostrazione
The existence of the $N\to\infty$ limiting density, $f_{X_\eta,v}$ is
a consequence of the result of Barra and Gaspard \cite{BG}. 
The proof is entirely analogous to the proof of theorem \ref{thm:2}.

We turn our attention to the $v\to\infty$ limit of this density. 
With $n$ a random variable uniformly distributed on the set 
$\{1,\ldots,N\}$ for some $N\in\N$,
the characteristic function for this random variable $X_\eta(n)$ is
\begin{equation*}
  e_{v,N}(\beta)\coloneq\E(\rme^{\rmi\beta X_{\eta}})=\frac{1}{N}
\sum_{n=1}^{N}f(k_n\vec{L})+\Ord(v\Delta L)
\end{equation*}
where $f:[0,\pi]^v\to\C$ is defined to be
\begin{equation*}
  f(\vec{x})\coloneq\exp\left(\frac{\rmi\beta}{v^2}\left(
\sum_{j=v+1}^{\alpha v}\sec^2x_j-\eta\sum_{i=1}^v \sec^2x_i\right)\right).
\end{equation*}
Following the argument of \cite{KMW} we can write
\begin{align*}
  \lim_{N\to\infty}&\frac{1}{N}\sum_{n=1}^{N}f(k_n\vec{L}) \\
&=\frac{1}{2\alpha v^2}\frac{1}{\pi^{\alpha v}}
\int_{-\infty}^{\infty}\int_0^{\pi}\!\cdots\!\int_0^{\pi}
\left(\sum_{j=1}^{\alpha v}\sec^2x_j\right)f(\vec{x})\exp\left(
\frac{\rmi\zeta}{v}\sum_{j=1}^{\alpha v}\tan x_j\right)\rmd^{\alpha v}\vec{x}
\rmd\zeta.
\end{align*}
Denoting this limit for each fixed $v, \beta$ by $e_v(\beta)$, we can write
\begin{equation}
  e_v(\beta)=\frac{1}{2\alpha v}\int_{-\infty}^{\infty}I_1 
I_2^{v-1} I_3^{\alpha v-v}
+(\alpha-1)I_4 I_2^v I_3^{\alpha v -v -1} \rmd\zeta
\label{eq:3:3}
\end{equation}
where the integrals $I_1,\ldots,I_4$ are:-
{\setlength\arraycolsep{2pt}
\begin{eqnarray}
  I_1&\coloneq& \frac{1}{\pi}\int_0^{\pi}\sec^2 x \exp\left(\frac{\rmi\zeta}{v}
\tan x-\frac{\rmi\beta\eta}{v^2}\sec^2 x\right)\rmd x, \label{eq:3:4}\\
 I_2&\coloneq& \frac{1}{\pi}\int_0^{\pi}\exp\left(\frac{\rmi\zeta}{v}
\tan x-\frac{\rmi\beta\eta}{v^2}\sec^2 x\right)\rmd x, \\
 I_3&\coloneq& \frac{1}{\pi}\int_0^{\pi}\exp\left(\frac{\rmi\zeta}{v}
\tan x+\frac{\rmi\beta}{v^2}\sec^2 x\right)\rmd x, \\
 I_4&\coloneq& \frac{1}{\pi}\int_0^{\pi}\sec^2 x \exp\left(\frac{\rmi\zeta}{v}
\tan x+\frac{\rmi\beta}{v^2}\sec^2 x\right)\rmd x. \label{eq:3:5}
\end{eqnarray}}
An integral similar to \eqref{eq:3:3} was tackled in \cite{KMW}. We quote
here the relevant results, {\it mutatis mutandis}.
Asymptotic analysis of the integrals in (\ref{eq:3:4}--\ref{eq:3:5}) gives
\begin{equation}
  I_1=\frac{v}{\sqrt{\pi\beta\eta}}\exp\left(-\frac{\rmi\pi}{4}+
\frac{\rmi\zeta^2}{4\beta\eta}\right)+
\Ord(1)\quad\mbox{as $v\to\infty$}
\end{equation}
and
\begin{equation}
  I_4=\frac{v}{\sqrt{\pi\beta}}\exp\left(\frac{\rmi\pi}{4}-\frac{\rmi\zeta^2}
{4\beta}\right)+\Ord(1)\quad\mbox{as $v\to\infty$}.
\end{equation}
For $I_2$ and $I_3$ we consider separately the cases $-\sqrt{v}<\zeta<\sqrt{v}$
and $|\zeta|>\sqrt{v}$.
For $-\sqrt{v}<\zeta<\sqrt{v}$,
\begin{equation}
  I_2^{v}=\exp\left(-\frac{2}{\sqrt{\pi}}\sqrt{\beta\eta}\exp\left(
\frac{\rmi\pi}{4}+\frac{\rmi\zeta^2}{4\beta\eta}\right)-
\zeta\erf\left(\frac{\rme^{-\rmi\pi/4}\zeta}{2\sqrt{\beta\eta}}\right)
\right)\left(1+\Ord\left(\frac{1+\zeta^2}{v}\right)\right)
\end{equation}
and
\begin{align}
  I_3^{(\alpha-1)v}=\exp\left(-\frac{2(\alpha-1)}{\sqrt{\pi}}\sqrt{\beta}
\exp\left(
-\frac{\rmi\pi}{4}-\frac{\rmi\zeta^2}{4\beta}\right)-
\zeta(\alpha-1)\right.\erf&\left.
\left(\frac{\rme^{\rmi\pi/4}\zeta}{2\sqrt{\beta}}\right)\right) \\
&\times\left(1+\Ord\left(\frac{1+\zeta^2}{v}\right)\right), \nonumber
\end{align}
both error estimates are as $v\to\infty$. For $|\zeta|>\sqrt{v}$,
\begin{equation}
  |I_2|\leq\frac{\sqrt{\beta\eta}}{v\pi}\left(\frac{\beta\eta}{v^2}
+\frac{\zeta^2}{\beta\eta}\right)^{-1}+\Ord(\zeta^{-3})
\end{equation}
as $\zeta\to\infty$, and
\begin{equation}
  |I_3|\leq\frac{\sqrt{\beta}}{v\pi}\left(\frac{\beta}{v^2}
+\frac{\zeta^2}{\beta}\right)^{-1}+\Ord(\zeta^{-3}).
\end{equation}
Using these estimates, we can find an expression for the limit of $e_v(\beta)$
as $v\to\infty$,
\begin{align*}
e(\beta)\coloneq\lim_{v\to\infty}e_v(\beta)&=
\frac{1}{2\alpha}\int_{-\infty}^{\infty}\frac{1}{\sqrt{\beta}}P_\eta\left(
\frac{\zeta}{\sqrt{\beta}}\right)\exp\left(-\sqrt{\beta}\tau_{\eta}
\left(\frac{\zeta}{\sqrt{\beta}}\right)\right)\rmd\zeta \\
&=\int_{-\infty}^{\infty}P_\eta(\xi)\exp\left(-\sqrt{\beta}\tau_{\eta}(\xi)
\right)\rmd\xi.
\end{align*}
For ease of notation, we have introduced
\begin{equation} \label{eq:def:P}
  P_\eta(\xi)\coloneq\frac{1}{\sqrt{\pi\eta}}\exp\left(\frac{-\rmi\pi}{4}+
\frac{\rmi\xi^2}{4\eta}\right)+\frac{(\alpha-1)}{\sqrt{\pi}}\exp
\left(\frac{\rmi\pi}{4}-\frac{\rmi\xi^2}{4}\right)
\end{equation}
and
\begin{align}
  \tau_\eta(\xi)\coloneq& \frac{2}{\sqrt{\pi}}\sqrt{\eta}\exp\left(
\frac{\rmi\pi}{4}+\frac{\rmi\xi^2}{4\eta}\right)+\xi\erf\left(
\frac{\rme^{-\rmi\pi/4}\xi}{2\sqrt{\eta}}\right) \nonumber \\
&+\frac{2(\alpha-1)}{\sqrt{\pi}}\exp\left(
-\frac{\rmi\pi}{4}-\frac{\rmi\xi^2}{4}\right)+\xi(\alpha-1)\erf\left(
\frac{\rme^{\rmi\pi/4}\xi}{2}\right). \label{eq:def:tau}
\end{align}
In the above, wherever $\sqrt{\beta}$ occurs for $\beta<0$, this should
be understood to mean $\pm\rmi\sqrt{-\beta}$ where the sign is taken 
in such a way that
\begin{equation*}
  e(-\beta)=\overline{e(\beta)},
\end{equation*}
the usual condition for the characteristic function of a probability 
density. This can always be done.
We also note that $e(0)=1$ which is consistent with $e(\beta)$ being the
characteristic function of a probability distribution.
$e(\beta)$ is continuous at $\beta=0$ since the defining integral is uniformly
convergent in $\beta$ (see lemma \ref{lem:app:3} below). 
Thus the limiting density, $f_{X_\eta}$ exists and is given by
\begin{equation} \label{eq:3:12}
  f_{X_\eta}(\sigma)=\frac{1}{2\pi}\int_{-\infty}^{\infty}\frac{1}{2\alpha}
\int_{-\infty}^{\infty}P_\eta(\xi)\exp(-\sqrt{\beta}\tau_{\eta}(\xi)-\rmi\sigma
\beta)\rmd\xi\rmd\beta,
\end{equation}
where we have made the substitution $\xi=\zeta/\sqrt{\beta}$.
We here switch the order of integration. This is a non-trivial operation 
since both integrals are improper. However in this case we can rigorously
justify the manoeuvre. Justification is provided in appendix \ref{app:a},
proposition \ref{prop:a}.
\begin{eqnarray}
  f_{X_\eta}(\sigma)&=&\frac{1}{2\pi\alpha}\Re\int_{-\infty}^{\infty}
P_\eta(\xi)\int_0^{\infty}\exp(-\sqrt{\beta}
\tau_\eta(\xi)-\rmi\sigma\beta)\rmd\beta\;\rmd\xi \nonumber\\
&=&\frac{1}{2\pi\alpha}\Re\int_{-\infty}^{\infty}
P_\eta(\xi)\left(\frac{1}{\rmi\sigma}-\frac{\sqrt{\pi}}{2}\frac{
\rme^{3\rmi\pi/4}\tau_\eta(\xi)}{(-\sigma)^{3/2}}
w\left(\frac{\rme^{3\rmi\pi/4}\tau_\eta(\xi)}{2\sqrt{-\sigma}}
\right)\right)\rmd\xi.\label{eq:3:10}
\end{eqnarray}
To evaluate the final integration we have used the following result 
(a variant of formula {\bf 3.462.5} in \cite{GR}),
\begin{equation}
  \int_0^{\infty} \exp(-ax-b\sqrt{x})\rmd x=\frac{1}{a}-\frac{\sqrt{\pi}}{2}
\frac{b}{a^{3/2}}w\left(\frac{\rmi b}{2a^{1/2}}\right).
\end{equation}
To conclude we observe that since 
\begin{equation*}
  \int_{-\infty}^{\infty} P_\eta(\xi)\rmd\xi=2\alpha\in\R
\end{equation*}
the first term in \eqref{eq:3:10} vanishes as it has zero real part.
\finire

Some properties of $w(z)$ are discussed in \cite{AS} chapter 7. We shall
use the following
\begin{lemma} \label{lem:w}
  The function $w(z)$ has the asymptotic expansion
  \begin{equation}
    w(z)\sim \frac{\rmi}{\sqrt{\pi}}\sum_{m=0}^{\infty} 
\frac{(2m)!}{4^m m! z^{2m+1}}
  \end{equation}
as $z\to\infty$, valid for $\displaystyle\frac{-\pi}{4}<\arg z<\frac{5\pi}{4}$.
\end{lemma}
\dimostrazione
This follows from the asymptotic expansion of $\erfc$,
\begin{equation*}
  \sqrt{\pi} z\rme^{z^2}\erfc(z)
\sim\sum_{m=0}^{\infty} \frac{(2m)!}{(4z^2)^m m!},
\end{equation*}
as $z\to\infty$, $|\arg z|<3\pi/4$,
taken from \cite{AS} (formula {\bf 7.1.23}; see also \cite{BH} exercise 3.11).
The series for $w$ comes from making the substitution $z\mapsto-\rmi z$.
\finire

Since $\erfc$ and $w$ are analytic functions, they are bounded in the
domain of validity of their asymptotic expansions quoted in lemma \ref{lem:w}.

\begin{figure}
  \begin{center}
\setlength{\unitlength}{7cm}
\begin{picture}(1,0.9)
\put(0.0,0.0){\includegraphics[angle=0,width=6.0cm,height=5cm]{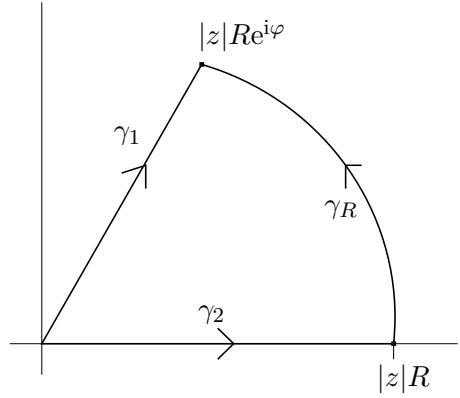}}
\put(0.2,0.45){$\gamma_1$}
\put(0.7,-0.02){$|z|R$}
\put(0.6,0.31){$\gamma_R$}
\put(0.36,0.11){$\gamma_2$}
\put(0.36,0.63){$|z|R\rme^{\rmi\varphi}$}
\end{picture}
\caption{The contours $\gamma_1$, $\gamma_2$ and $\gamma_R$.}
\label{fig:2}
\end{center}
\end{figure}

\begin{proposition} \label{prop:w}
  Let $z\in\C$ and $-\pi/4<\arg z<5\pi/4$. Then
  \begin{equation}
    \int_0^R zw(zp)\rmd p=\frac{\sqrt{\pi}}{2}-\frac{\arg(z)}{\sqrt{\pi}}
+\frac{\rmi}{\sqrt{\pi}}\log|z|+{\rmi}C_R+
\Ord\left(\frac{1}{|z|^2 R^2}\right)
\label{eq:wint}
  \end{equation}
where $C_R\in\R$ is independent of $z$, but may depend on $R$.
\end{proposition}
\dimostrazione
Write $z=|z|\rme^{\rmi\varphi}$ where
$\varphi=\arg z$. Then
\begin{equation}
  \int_{0}^{R} zw(zp)\rmd p=\int_0^{|z|R} 
\rme^{\rmi\varphi}w(\rme^{\rmi\varphi}p)\rmd p
\end{equation}
via $p\mapsto p/|z|$. Using Cauchy's theorem,
\begin{equation}
  \int_0^{|z|R} 
\rme^{\rmi\varphi}w(\rme^{\rmi\varphi}p)\rmd p=\int_{\gamma_1} w(t)\rmd t
=\int_{\gamma_2} w(t)\rmd t+\int_{\gamma_R} w(t)\rmd t.
\end{equation}
The contours $\gamma_1, \gamma_2$ and $\gamma_R$ in the complex $t$-plane
are illustrated in figure \ref{fig:2}.

On $\gamma_2$,
\begin{align*}
  \int_{\gamma_2} w(t)\rmd t =& \int_0^{|z|R} w(x)\rmd x\\
=& \int_0^1 w(x)\rmd x +\int_1^{|z|R} \left(w(x)-\frac\rmi{\sqrt{\pi}x}\right)
 \rmd x
+ \int_1^{|z|R} \frac\rmi{\sqrt{\pi}x}\rmd x\\
=& \int_0^1 w(x)\rmd x +\int_1^\infty \left(w(x)-\frac\rmi{\sqrt{\pi}x}\right)
 \rmd x\\
&- \int_{|z|R}^\infty \left(w(x)-\frac\rmi{\sqrt{\pi}x}\right) \rmd x+
\int_1^{|z|R} \frac\rmi{\sqrt{\pi}x}\rmd x \\
=& \int_0^{\infty} \rme^{-x^2}\rmd x +\rmi\times\mbox{const}
- \int_{|z|R}^\infty \left(w(x)-\frac\rmi{\sqrt{\pi}x}\right) \rmd x+ 
\int_1^{|z|R} \frac\rmi{\sqrt{\pi}x}\rmd x,
\end{align*}
using $w(z)=\rme^{-x^2}(1+\rmi\erfi z)$ to separate the real
and imaginary contributions. The imaginary part goes into the constant
$C_R$, the value of which is not important since we shall always be
considering only the real part of resulting expressions. 
By the use of lemma \ref{lem:w},
\begin{equation} \label{potato:1}
  \int_{\gamma_2} w(t)\rmd t=\frac{\sqrt{\pi}}{2}+\rmi\times\mbox{const}
+\frac{\rmi}{\sqrt{\pi}}(\log R+\log|z|)+\Ord\left(\frac{1}{|z|^2R^2} \right)
\end{equation}
as $R\to\infty$ uniformly for $|z|\geq c$ for some $c$.

On $\gamma_R$,
\begin{eqnarray}
  \int_{\gamma_R} f(t)\rmd t &=& \int_0^\varphi \rmi |z|R \rme^{\rmi \theta}
w(|z|R\rme^{\rmi\theta})\rmd\theta \nonumber \\
&=&\int_0^\varphi \rmi |z|R \rme^{\rmi\theta}\left(
\frac{-1}{\sqrt{\pi}\rmi |z|R\rme^{\rmi\theta}}+\Ord\left(
\frac{1}{|z|^3R^3}\right)\right)\quad
\mbox{by lemma \ref{lem:w}}\nonumber \\
&=&\frac{-\varphi}{\sqrt{\pi}}+\Ord\left(\frac{1}{|z|^2 R^2}\right).
\label{potato:2}
\end{eqnarray}
Combining \eqref{potato:1} and \eqref{potato:2} gives \eqref{eq:wint}.
\finire

The following lemma from probability theory will also be useful.
\begin{lemma} \label{lem:3:1}
  Let $U, V$ be random variables, then
  \begin{equation*}
    \Prob\left(\frac{U}{V}<\eta\right)=\int_{-\infty}^0 f_{X_{\eta}}(\sigma)
\rmd\sigma
  \end{equation*}
where $f_{X_\eta}$ is the probability density function of the random variable
$X_\eta\coloneq U-\eta V$.
\end{lemma}
\dimostrazione
This follows immediately from the fact that
\begin{equation*}
  \Prob\left(\frac{U}{V}<\eta\right)=\Prob(U-\eta V<0).
\end{equation*}
\finire

\section{Proof of theorem \ref{thm:3}} \label{sec:6}
We first observe that
\begin{equation} \frac{\sum_{i=1}^v L_i\sec^2{k_nL_i}}{\sum_{j=1}^{\alpha v}L_j\sec^2 k_nL_j} = 
\left(1+\frac{\sum_{j=v+1}^{\alpha v} L_j\sec^2{k_nL_j}}{\sum_{1=1}^{v}
L_i\sec^2 k_nL_i}\right)^{-1},
\end{equation}
so we can concentrate on finding the probability distribution, 
$\tilde{F}(\eta)$, of
\begin{equation*}
\frac{\sum_{j=v+1}^{\alpha v} L_j\sec^2{k_nL_j}}{\sum_{1=1}^{v}
L_i\sec^2 k_nL_i}
\end{equation*}
as $v\to\infty$. This will be then related to the distribution in which we
are interested by a simple transformation (see equation 
\eqref{eq:last} below).
In light of lemma \ref{lem:3:1}, $\tilde{F}(\eta)$ is given by
\begin{equation*}
  \tilde{F}(\eta)=\int_{-\infty}^{0} f_{X_\eta}(\sigma)\rmd\sigma 
\end{equation*}
with $f_{X_\eta}$ found in proposition \ref{prop:5:2}.
It is more instructive to take the range of integration 
from $-R^2$ to $0$, with a view
to taking the limit $R\to\infty$ later,
\begin{equation}  \label{eq:greg:greg}
  \int_{-R^2}^0 f_{X_\eta}(\sigma)\rmd\sigma=
\frac{-1}{\sqrt{\pi}\alpha}\Re\int_{-\infty}^{\infty} \int_{-R^2}^0
P_{\eta}(\xi)\frac{\rme^{3\pi\rmi/4}\tau_{\eta}(\xi)}{4(-\sigma)^{3/2}}
w\left( \frac{\rme^{3\pi\rmi/4}\tau_{\eta}(\xi)}{\sqrt{-\sigma}}\right)
\rmd\sigma\rmd\xi. 
\end{equation}
The $\xi$-integral is uniformly convergent by lemma \ref{lem:app:6}, so
we have legitimately switched the order of integration.
We can then write
\begin{align*}
  \int_{-R^2}^0 f_{X_\eta}(\sigma)\rmd\sigma=&
\frac{-1}{\alpha\sqrt{\pi}}\Re\int_{-\infty}^{\infty} \int_{-\infty}^0
P_{\eta}(\xi)\frac{\rme^{3\pi\rmi/4}\tau_{\eta}(\xi)}{4(-\sigma)^{3/2}}
w\left( \frac{\rme^{3\pi\rmi/4}\tau_{\eta}(\xi)}{\sqrt{-\sigma}}\right)
\rmd\sigma\rmd\xi\\
&-\frac{-1}{\sqrt{\pi}\alpha}\Re\int_{-\infty}^{\infty} \int_{-\infty}^{-R^2}
P_{\eta}(\xi)\frac{\rme^{3\pi\rmi/4}\tau_{\eta}(\xi)}{4(-\sigma)^{3/2}}
w\left( \frac{\rme^{3\pi\rmi/4}\tau_{\eta}(\xi)}{\sqrt{-\sigma}}\right)
\rmd\sigma\rmd\xi.
\end{align*}
The second term vanishes as $R\to\infty$, as shown in proposition
\ref{prop:app:2}.
Making the substitution $(-\sigma)^{-1/2}=2p$, leads to
\begin{equation*}
  \tilde{F}(\eta)=\frac{-1}{\sqrt{\pi}{\alpha}}\Re\int_{-\infty}^{\infty}
\int_0^{\infty} P_\eta(\xi)\rme^{3\pi\rmi/4}\tau_{\eta}(\xi)
w(\rme^{3\pi\rmi/4}\tau_{\eta}(\xi)p)\rmd p\rmd\xi.
\end{equation*}
We can apply proposition \ref{prop:w} with 
$z=\rme^{3\rmi\pi/4}\tau_\eta(\xi)$. 
We can integrate $\xi$ out of the error term provided by this proposition
since
\begin{equation*}
  \int_{-\infty}^{\infty} \frac{1}{\tau_\eta(\xi)^2}\rmd\xi<\infty.
\end{equation*}
We get, finally,
\begin{eqnarray}
\tilde{F}(\eta)&=&\frac{-1}{\alpha\sqrt{\pi}}
\Re\int_{-\infty}^{\infty} P_\eta(\xi)
\left(\frac{\sqrt{\pi}}{2}-\frac{\arg(\rme^{3\rmi\pi/4}\tau_\eta(\xi))}
{\sqrt{\pi}}+\rmi\frac{\log|\tau_\eta(\xi)|}{\sqrt{\pi}}\right)\rmd\xi
\nonumber\\
&=&\frac{1}{2}+\frac{1}{\pi\alpha}\Re\int_{-\infty}^{\infty} P_\eta(\xi)
\left(\arg(\tau_\eta(\xi))-\rmi\log|\tau_{\eta}(\xi)|\right)\rmd\xi.
\end{eqnarray}
This is related to $F(\eta)$ by
\begin{equation}
  F(\eta)=1-\tilde{F}\left(\frac{1}{\eta}-1\right).
\label{eq:last}
\end{equation}
\finire
%
%

\section{Scarred states on finite star graphs} \label{sec:7}

We recall that theorem \ref{thm:1} is concerned with the quantity 
$\langle\boldpsi^{(n)}|\vec{f}|\boldpsi^{(n)}\rangle$ which can be
written as 
\begin{equation*}
  \langle\boldpsi^{(n)}|\vec{f}|\boldpsi^{(n)}\rangle=
  \sum_{i=1}^{v} \int_0^{L_i} |\psi_i^{(n)}(x)|^2f_i(x)\rmd x.
\end{equation*}
Now using (\ref{eq:efun}) and the identity $\cos^2\theta=
1/2+(1/2)\cos 2\theta$ yields
\begin{equation}
  \label{eq:q_expansion}
  \langle \boldpsi^{(n)} | \vec{f} | \boldpsi^{(n)} \rangle
  = \sum_{i=1}^{v} \frac{{A_i^{(n)}}^2}{2}
  \left(\int_0^{L_i} f_i(x)\rmd x + \int_0^{L_i}
    \cos2k_n(x-L_i)f_i(x)\rmd x
  \right).
\end{equation}
Since we are interested in a subsequence $k_{n_r} \to \infty$, we
may hope that the second integrals do not survive.  Thus our prime
concern are the prefactors
\begin{equation*}
  {A_i^{(n)}}^2 
  = \frac{2\sec^2{k_nL_i}}{\sum_{j=1}^{v} L_j\sec^2{k_nL_j}}.
\end{equation*}
and to prove the theorem we need to find $k_n$ such
that the prefactors are small for all $i$ other than (without loss of
generality) $1$ and $2$.

To do this we study the poles of the function $Z(k,\vec{L})$.  Let
\begin{equation*}
  p_{n,i}\coloneq\frac{\pi}{L_i}\left(\frac{1}{2}+n\right).
\end{equation*}
Then $|\tan pL_i|\to\infty$ and $\sec pL_i\to\infty$ as $p\to p_{n,i}$.

\begin{figure}
  \begin{center}
    \setlength{\unitlength}{6cm}
    \begin{picture}(2,0.8)
      \put(-0.1,0.1)
      {\includegraphics[angle=0,scale=0.65]{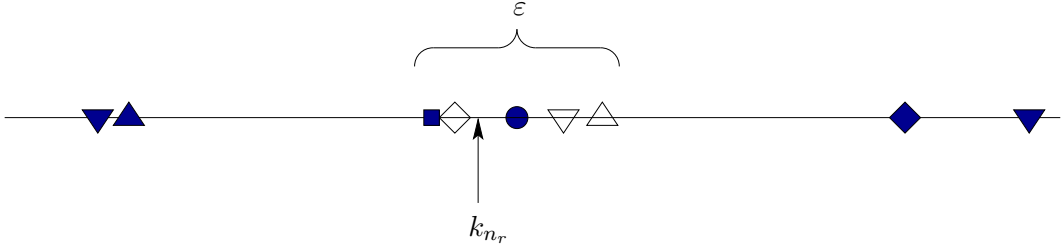}}
      \put(0.93,0.03){$k_{n_r}$}
      \put(1.03,0.52){$\epsilon$}
    \end{picture}
    \caption{Poles $p_{n,i}$ and nodes $\ell_{n,i}$ on the real line.
      Different symbols correspond to different values of $i$ (the
      circle corresponds to $p_{n,1}$).  Filled symbols correspond to
      the poles, empty symbols to the nodes.  In this example $v=5$ and
      $v^{\ast}=2$.}
    \label{fig:poles}
  \end{center}
\end{figure}

Since the function $Z(k,\vec{L})$ is an increasing function of $k$,
between any two poles there is a zero.  We will use this important
feature to ``trap'' zeros $k_{n_r}$ of $Z(k,\vec{L})$ between pairs of
nearby poles $p_{n,1}$ and $p_{m,2}$, also requiring that all other
poles are far away (see Fig.~\ref{fig:poles}).  The implications will
be that as $r\to\infty$,
\begin{equation*}
  \sec^2{k_{n_r}L_i} \gg \sec^2{k_{n_r}L_j} \qquad \mbox{ with } i=1,2 
  \mbox{ and } j > 2,
\end{equation*}
and
\begin{equation*}
  \sec^2 k_{n_r}L_1\sim \sec^2 k_{n_r}L_2,
\end{equation*}
ensuring that (\ref{eq:scars_thm}) holds.

To make the above arguments rigorous we need the following propositions.

\begin{proposition} \label{prop:1}
  Let the elements of $\vec{L}$ be linearly independent over
  $\Q$.  Let $1<v^{\ast}< v$ for $v\geq 3$.  Given $\epsilon>0$ there
  exist infinitely many $n\in\N$ such that
  \begin{itemize}
  \item for each $i=2,\ldots,v^{\ast}$ there exists $m\in\N$
    satisfying
    \begin{equation}
      \label{eq:item1}
      |p_{m, i} - p_{n, 1}| \leq \epsilon/2
    \end{equation}
  \item for all $i=v^{\ast}+1,\ldots,v$ and for all $m\in\N$
    \begin{equation}
      \label{eq:item2}
      |p_{m, i} - p_{n, 1}| \geq \frac{\pi}{2L_i} - \epsilon/2.
    \end{equation}
  \end{itemize}
\end{proposition}
\dimostrazione 
The idea behind the proof is that for linearly independent elements of
$\vec{L}$ the poles $p_{\cdot, i}$ for different $i$ behave like
independent random variables, therefore every permitted pole
configuration happens infinitely often.  To substantiate this claim we
express the nearest-pole distances as the states of an ergodic dynamical
system.

For $n\in\N$ and $i=2,\ldots, v^{\ast}$, let $\delta_{n,i}$ denote
the distance between $p_{n,1}$ and the closest pole of $\tan kL_i$;
\begin{equation*}
  \delta_{n,i} \coloneq p_{n,1} - p_{m,i},
\end{equation*}
where $m$ is such that 
\begin{equation*}
  |p_{m,i}-p_{n,1}|=\min_{m}\{ |p_{m,i}-p_{n,1}|\}.
\end{equation*}
Since the poles of $\tan kL_i$ are $\pi/L_i$-periodic, we have
\begin{eqnarray}
  \delta_{n,i} + \frac{\pi}{2L_i}
  &=& p_{n, 1} - p_{0, i} + \frac{\pi}{2L_i} \mod{\frac{\pi}{L_i}} \\
  &=& \frac{\pi}{2L_1} 
  + \frac{\pi}{L_1}n \mod{\frac{\pi}{L_i}}.
  \label{eqn:1:1}
\end{eqnarray}
Let $\ell_{m,i}$ denote the $m^{\rm th}$ zero of $\tan kL_i$.  We
note that (\ref{eq:item2}) is implied by the condition that
\begin{equation*}
  |\ell_{m, i} - p_{n, 1}| \leq \epsilon/2    
\end{equation*}
for some $m\in\N$.  For $i=v^{\ast}+1,\ldots,v$, define $\eta_{n,i}$
to be the distance between $p_{n,1}$ and the closest zero of $\tan
kL_i$.  Similarly to (\ref{eqn:1:1})
\begin{equation}
  \label{eqn:1:2}
  \eta_{n,i}  + \frac{\pi}{2L_i} = 
\frac{\pi}{2}\left( \frac{1}{L_1} + \frac{1}{L_i} \right)
+ \frac{\pi}{L_1}n \mod{\frac{\pi}{L_i}}.
\end{equation}

From (\ref{eqn:1:1}) and (\ref{eqn:1:2}), $\delta_{n,i}$ and $\eta_{n,i}$
satisfy the recurrence
\begin{equation}
  \left\{
    \begin{array}{rcll}
      \delta_{n+1,i} & = &
      \delta_{n,i}+\frac{\pi}{L_1}\mod{\frac{\pi}{L_i}} & 
      \quad \mbox{$i=2,\ldots,v^{\ast}$}\\ 
      \eta_{n+1,i} & = &
      \eta_{n,i}+\frac{\pi}{L_1}\mod{\frac{\pi}{L_i}} & 
      \quad \mbox{$i=v^{\ast}+1,\ldots,v$}
    \end{array}\right.
  \label{eq:1:3}
\end{equation}
Since the bond lengths are not rationally related, the dynamical
system (\ref{eq:1:3}) is equivalent to an irrational translation on a
torus. In this case, Weyl's equidistribution result \cite{W} applies,
and any subset of the torus with positive Lebesgue measure is visited
infinitely many times. The volume of the area in $\delta-\eta$ space
defined by
\begin{equation}
  \label{eq:def_area}
  -\epsilon/2 < \delta_{n,i}, \eta_{n,i} < \epsilon/2
\end{equation}
is non-zero and so there are infinitely many $n$ for which
(\ref{eq:def_area}), and therefore (\ref{eq:item1}-\ref{eq:item2}),
are satisfied.
\finire

The interpretation of proposition \ref{prop:1} is that we can find situations
on the real line where $v^{\ast}$ poles of the functions $\tan{kL_i}$ are
bunched together and the remaining $v-v^{\ast}$ poles are not close
to these bunched poles (see figure \ref{fig:poles}).

\begin{proposition} \label{prop:2}
  Under the conditions of proposition \ref{prop:1} there is a
  subsequence $(k_{n_r})\subseteq(k_n)$ for which
  \begin{eqnarray*}
    \sec^2{k_{n_r}L_i}&\to& \infty \qquad \mbox{for }
    i=1,\ldots,v^{\ast}\\
    \sec^2{k_{n_r}L_i}&\to& 1 \qquad \mbox{for }
    i=v^{\ast}+1,\ldots,v
  \end{eqnarray*}
  as $r\to\infty$.
\end{proposition}
\dimostrazione
Let $(\epsilon_r)$ be a sequence satisfying $\epsilon_r\to0$ as $r\to\infty$.
We choose $k_{n_r}$ as follows. 

Applying proposition \ref{prop:1} with $\epsilon=\epsilon_r$ yields a
set of $v^{\ast}$ poles of $Z(k,\vec{L})$ inside a region with width
$\epsilon_r$.  Since there is a zero of $Z(k,\vec{L})$ between any two
poles of $Z(k,\vec{L})$, we can find $v^{\ast}-1$ zeros in this
region.  Set $k_{n_r}$ to be one of these zeros.

From proposition \ref{prop:1} we have
\begin{eqnarray*}
  |k_{n_r}-p_{m,i}| &\leq& \epsilon_{r} \qquad \mbox{for all }
  i= 1, \ldots, v^{\ast} \mbox{ and some } m = m(r,i)\\
  |k_{n_r}-\ell_{m,i}| &\leq& \epsilon_{r} \qquad \mbox{for all }
  i= v^{\ast}+1, \ldots, v \mbox{ and some } m = m(r,i).
\end{eqnarray*}
Since $\sec^2 L_ip_{m,i} = \infty $, $\sec^2 L_i\ell_{m,i} = 1$ and
$\sec \theta$ is a periodic function, the statement of the proposition
follows trivially.  \finire

\begin{corollary} \label{cor:3}
  If $v^{\ast}=2$ in proposition \ref{prop:2} then we additionally have
  \begin{equation*}
    \lim_{r\to\infty}\frac{\sec^2{k_{n_r}L_1}}{\sec^2{k_{n_r}L_2}}=1.
  \end{equation*}
\end{corollary}
\dimostrazione
We recall that since $k_{n_r}$ is an eigenvalue, $Z(k_{n_r},\vec{L})=0$, 
and hence
\begin{equation}
  \label{eq:k_n_a_root}
  \tan{k_{n_r}L_1} = - \tan{k_{n_r}L_2} - \tan{k_{n_r}L_3}
  - \cdots - \tan{k_{n_r}L_v}.
\end{equation}
On the other hand, by proposition \ref{prop:2},
$\tan k_{n_r}L_i$ remains bounded for $i>2$ and tends to
infinity for $i=1,2$.  Dividing (\ref{eq:k_n_a_root}) through by
$\tan{k_{n_r}L_2}$ we obtain
\begin{equation*}
  \lim_{r\to\infty} \frac{ \tan{k_{n_r}L_1}}{\tan{k_{n_r}L_2}} = -1.
\end{equation*}
Further observations that $\sin^2{k_{n_r}L_i} \to 1$ for $i=1,2$ and
$\sec^2 \theta = \tan^2\theta / \sin^2 \theta$ conclude the proof.
\finire

\begin{lemma}\label{lem:4}
  Let $f:[0,L]\to\R$ be continuously differentiable.  Then
  \begin{equation*}
    \lim_{k\to\infty}\int_0^L \cos (kx) f(x)\rmd x = 0.
  \end{equation*}
\end{lemma}
\dimostrazione
Integration by parts yields
\begin{equation*}
  \int_0^L \cos (kx) f(x)\rmd x = \frac{1}{k} \left(\sin(kL)f(L) 
    - \int_0^L \sin (kx) f'(x)\rmd x \right)
\end{equation*}
and the statement follows immediately from the boundedness of $f$ and
its derivative. 
\finire

\noindent{\sl Proof of theorem \ref{thm:1}.}\phantom{X}
Without loss of generality, we can assume that $i_1=1$ and $i_2=2$. We take
the subsequence whose existence is guaranteed by proposition \ref{prop:2}
with $v^{\ast}=2$. By corollary \ref{cor:3},
\begin{equation*}
  \lim_{r\to\infty} {A_i^{(n_r)}}^2 
  = 2 \left( \sum_{j=1}^v 
    L_j \lim_{r\to\infty}\frac{\sec^2{k_{n_r}L_j}}{\sec^2{k_{n_r}L_i}}
  \right)^{-1}
  = \left\{ 
    \begin{array}{lr}
      2(L_1 + L_2)^{-1} & \mbox{ if }i = 1, 2\\
      0 & \mbox{ otherwise}.
    \end{array}
  \right.
\end{equation*}
We use lemma \ref{lem:4} to get rid of the second integrals in
(\ref{eq:q_expansion}) and conclude
\begin{equation*}
  \lim_{r\to\infty} \langle \boldpsi^{(n_r)} | \vec{f} |
  \boldpsi^{(n_r)} \rangle 
  = \frac{1}{L_1+L_2} \left(\int_0^{L_1} f_1(x)\rmd x 
    + \int_0^{L_2} f_2(x)\rmd x\right).
\end{equation*}
\finire

\bigskip
\bigskip
 
{\bf Acknowledgements:} 
GB is grateful to the University of Bristol 
for hospitality during visits while part of this research was carried
out. 
BW wishes to thank the University of 
Strathclyde for hospitality. 

BW has been financially supported by an EPSRC studentship 
(Award Number 0080052X). 

We gratefully acknowledge the support of the
European Commission under the Research Training Network (Mathematical Aspects
of Quantum Chaos) HPRN-CT-2000-00103 of the IHP Programme.

\appendix
\section{Appendix: The order of integration in \eqref{eq:3:12}} \label{app:a}
In this appendix we deal with some technical issues regarding the 
exchange of order of integration in \eqref{eq:3:12}.

We first consider some asymptotics of $\tau_\eta(\xi)$.
\begin{lemma} \label{lem:app:1}
  For $\xi\in\R$,
  \begin{equation*}
    \tau_\eta(\xi)=\alpha|\xi|+\Ord_{\eta}(\xi^{-2})
  \end{equation*}
as $|\xi|\to\infty$, where the error estimate depends on $\eta$.
\end{lemma}
\dimostrazione
We first note that $\tau_\eta$ is an even function, so we may assume
$\xi>0$, and the result for $\xi<0$ will follow by symmetry.
We can write $\tau_\eta(\xi)$ as
\begin{equation*}
  \tau_\eta(\xi)=\sqrt{\eta} \overline{t\left(\frac{\xi}{\sqrt{\eta}}\right) }
+(\alpha-1)t(\xi)
\end{equation*}
where 
\begin{equation} \label{eq:zero}
  t(\xi)\coloneq \frac{2}{\sqrt{\pi}}\exp\left( -\frac{\rmi\pi}{4}-
\frac{\rmi \xi^2}{4}\right)+\xi\erf\left(\frac{\rme^{\rmi\pi/4}\xi}{2}\right).
\end{equation}
We expand the error function asymptotically,
\begin{align}
  \erf\left( \frac{\rme^{\rmi\pi/4}\xi}{2} \right) &= 
  1 -  \erfc\left( \frac{\rme^{\rmi\pi/4}\xi}{2} \right)  \nonumber \\
  &= 1-\frac{2}{\xi\sqrt{\pi}}  \label{eq:potato}
\exp\left(-\frac{\rmi\xi^2}{4}-\frac{\rmi\pi}{4}\right)+\Ord(\xi^{-3}) 
\end{align}
as $\xi\to\infty$. Substituting \eqref{eq:potato} into \eqref{eq:zero}
gives
\begin{equation}
  t(\xi)=\xi+\Ord(\xi^{-2}), \qquad\mbox{as $\xi\to\infty$,}
\end{equation}
and the lemma follows.
\finire

\begin{lemma} \label{lem:app:2}
For $\xi>0$, $\displaystyle \Re\;\frac{\rmd\tau_\eta}{\rmd\xi} \geq0$ and 
for all $\xi\in\R$, there exists $\tau^\ast>0$ such that
\begin{equation*}
  \Re\;\tau_\eta(\xi)\geq \tau^\ast.
\end{equation*}
\end{lemma}
\dimostrazione
By differentiation,
\begin{align*}
  \frac{\rmd t}{\rmd \xi}&=\erf\left( \frac{\rme^{\rmi\pi/4}\xi}{2}  \right)\\
&=\frac{2}{\sqrt{\pi}}\rme^{\rmi\pi/4}\int_0^{\xi/2}
\rme^{-\rmi r^2}\rmd r.
\end{align*}
We see that
\begin{equation*}
  \Re\;\frac{\rmd t}{\rmd \xi}=\frac{2}{\sqrt{\pi}}\int_0^{\xi/2}
\cos\left(r^2-\frac{\pi}{4}\right)\rmd r \geq 0.
\end{equation*}
Thus,
\begin{equation*}
  \Re\;\frac{\rmd\tau_{\eta}}{\rmd\xi}=\Re\; t'\left(\frac{\xi}{\sqrt{\eta}}
\right)+(\alpha-1)\Re\;t'(\xi) \geq 0.
\end{equation*}
Hence it follows that $\Re\: \tau_{\eta}(\xi)\geq \Re\: \tau_\eta(0)
=\sqrt{2}(\sqrt{\eta}+\alpha-1)/\sqrt{\pi}\eqcolon\tau^\ast$.
\finire

\begin{lemma}   \label{lem:app:3}
  The integral
  \begin{equation*}
    \int_0^{\infty} P_{\eta}(\xi)\exp(-\sqrt{\beta}\tau_\eta(\xi))\rmd\xi
  \end{equation*}
is uniformly convergent for $\beta\in[0,\beta_0]$ for all $\beta_0>0$.
\end{lemma}
\dimostrazione
By making the substitution $\nu=\xi^2$ we can consider the uniform convergence
of 
\begin{equation*}
  \int^{\infty} P_\eta(\sqrt{\nu})\exp(-\sqrt{\beta}\tau_\eta(\sqrt{\nu}))
\frac{\rmd\nu}{\sqrt{\nu}}.
\end{equation*}
Let
\begin{align*}
  f(\nu,\beta)&\coloneq \frac{1}{\nu^{1/4}}\exp(-\sqrt{\beta}\Re\:\tau_\eta
(\sqrt{\nu})),\\
\phi(\nu,\beta)&\coloneq \frac{P_{\eta}(\sqrt{\nu})}{\nu^{1/4}}\exp
(-\sqrt{\beta}\rmi\Im\:\tau_\eta(\sqrt{\nu})).
\end{align*}
By lemma \ref{lem:app:1}, $\Im\; \tau_{\eta}(\sqrt{\nu})=\Ord(\nu^{-1})$ as
$\nu\to\infty$ (We drop the $\eta$-dependence since we are concerned here only
with fixed $\eta$). So
\begin{equation}
  \exp(-\sqrt{\beta}\rmi \Im\; \tau_{\eta}(\sqrt{\nu}))=1+\Ord(\nu^{-1})
\end{equation}
uniformly for $\beta\in[0,\beta_0]$. This means that
\begin{equation*}
  \int^{\infty} \phi(\nu,\beta)\rmd\nu
\end{equation*}
converges uniformly. i.e.\ given any $\epsilon>0$ there exists $\nu_1$ such
that for any $\nu_2>\nu_1$,
\begin{equation*}
  \left| \int_{\nu_1}^{\nu_2} \phi(\nu,\beta)\rmd\nu \right|<\epsilon
\end{equation*}
for all $\beta\in[0,\beta_0]$. $f(\nu,\beta)$ is differentiable
in $\nu$, and decreasing, so that
\begin{equation*}
  \frac{\partial f}{\partial \nu}\leq0.
\end{equation*}
If we let $\psi(\nu,\beta)\coloneq\int_{\nu_1}^{\nu}\phi(\nu',\beta)\rmd\nu'$
then integrating by parts gives
\begin{align*}
  \int_{\nu_1}^{\nu_2} f(\nu,\beta)\phi(\nu,\beta)\rmd\nu &=
f(\nu_2,\beta)\psi(\nu_2,\beta)-\int_{\nu_1}^{\nu_2} \frac{\partial f}
{\partial \nu}(\nu,\beta)\psi(\nu,\beta)\rmd\nu \\
&\leq f(\nu_2,\beta)\epsilon -\epsilon\int_{\nu_1}^{\nu_2} \frac{\partial f}
{\partial \nu}\rmd\nu\\
&=\epsilon f(\nu_1,\beta),
\end{align*}
where we have used the mean value theorem for integrals. If additionally,
$\nu_1>1$ then $f(\nu_1,\beta)<1$ and we are done. 
\finire

\begin{corollary} \label{corr:app:1}
  \begin{equation}
    \int_0^{\infty}\int_0^1 P_\eta(\xi)\rme^{-\sqrt{\beta}\tau_\eta(\xi)-
\rmi\sigma\beta}\rmd\beta\rmd\xi =
 \int_0^{1}\int_0^{\infty} P_\eta(\xi)\rme^{-\sqrt{\beta}\tau_\eta(\xi)-
\rmi\sigma\beta}\rmd\xi\rmd\beta
  \end{equation}
\end{corollary}
\dimostrazione
This follows immediately from lemma \ref{lem:app:3}, see, for example,
\S11.55.II of \cite{S}.
\finire

\begin{lemma} \label{lem:app:4}
The integral
\begin{equation*}
  \int_1^{\infty} \exp(-\sqrt{\beta} \tau_\eta(\xi)-\rmi\sigma\beta)\rmd\beta
\end{equation*}
is uniformly convergent for $\xi\in[0,\xi_0]$ for all $\xi_0>0$.
\end{lemma}
\dimostrazione
We, in fact, prove the stronger statement that the integral in question
is uniformly convergent for all $\xi>0$.
Taking $M(\beta)\coloneq\rme^{-\tau^{\ast}\sqrt{\beta}}$,
\begin{equation*}
  |\exp(-\sqrt{\beta} \tau_\eta(\xi)-\rmi\sigma\beta) |\leq M(\beta)
\end{equation*}
and the integral is uniformly convergent by the Weierstrass $M$-test.
\finire

\begin{lemma} \label{lem:app:5}
  The iterated integral
  \begin{equation}
    \int_0^{\infty} \int_1^{R^2} P_\eta(\xi)\rme^{-\sqrt{\beta}\tau_\eta(\xi)-
\rmi\sigma\beta}\rmd\beta\rmd\xi
  \end{equation}
converges uniformly for $R>1$.
\end{lemma}
\dimostrazione
We shall first consider the case where $\sigma<0$. A lengthy calculation
gives
\begin{align*}
  \int_1^{R^2} \rme^{-\sqrt{\beta}\tau_\eta(\xi)-
\rmi\sigma\beta}\rmd\beta = & \frac{1}{\rmi\sigma}\rme^{-\rmi\sigma
-\tau_{\eta}(\xi)}
-\frac{1}{\rmi\sigma}\rme^{-\rmi R^2\sigma-R\tau_{\eta}(\xi)} \\
& -\frac{\sqrt{\pi}\tau_{\eta}(\xi)}{2\rme^{-3\pi\rmi/4}(-\sigma)^{3/2}}
\exp\left(\frac{\tau_{\eta}(\xi)^2}{4\rmi\sigma}\right)\left[
\erfc\left(\rme^{-\rmi\pi/4}\sqrt{-\sigma}+\frac{\tau_\eta(\xi)\rme^{\rmi\pi/4}}
{2\sqrt{-\sigma}}\right)\right. \\
& \quad - \left.\erfc\left( R\rme^{-\rmi\pi/4}\sqrt{-\sigma} +
\frac{\tau_\eta(\xi)\rme^{\rmi\pi/4}}{2\sqrt{-\sigma}}\right)\right]. 
\end{align*}
By Lemma \ref{lem:app:1}, $\tau_\eta(\xi)\sim \alpha\xi$ as $\xi\to\infty$, so
\begin{equation*}
  \int_0^{\infty} P_\eta(\xi)\rme^{-\rmi R^2\sigma-R\tau_{\eta}(\xi)} \rmd\xi
\end{equation*}
is uniformly convergent for $R>1$ by the Wierstrass $M$-test with
$M(\xi)\coloneq C\rme^{-\tau_{\eta}(\xi)}$ for some constant $C$ which does not
depend on $\xi$.

We can write
\begin{align*}
  \exp\left(\frac{-\rmi\tau_{\eta}(\xi)^2}{4\sigma}\right)
&\erfc\left(R\rme^{-\rmi\pi/4}\sqrt{-\sigma}+
\frac{\tau_\eta(\xi)\rme^{\rmi\pi/4}}{2\sqrt{-\sigma}}\right)\\
&=
\exp\left(-R^2\rmi\sigma -R\tau_\eta(\xi)\right)w\left(
R\rme^{\rmi\pi/4}\sqrt{-\sigma}+\frac{\tau_\eta(\xi)\rme^{3\pi\rmi/4}}
{2\sqrt{-\sigma}}\right)
\end{align*}
and since $w(z)=\Ord(z^{-1})$ as $z\to\infty$ and $|\tau_\eta(\xi)|\geq
\tau^{\ast}$, 
\begin{equation*}
  w\left(
R\rme^{\rmi\pi/4}\sqrt{-\sigma}+\frac{\tau_\eta(\xi)\rme^{3\pi\rmi/4}}
{2\sqrt{-\sigma}}\right)=\Ord(1)
\end{equation*}
as $\xi\to\infty$, uniformly for $R>1$. Since 
$|\exp(-R\tau_\eta(\xi))|\leq\exp(-\tau_\eta(\xi))$
we see that the convergence of
\begin{equation*}
\int_0^{\infty} P_\eta(\xi)\tau_\eta(\xi) \exp\left(\frac{-\rmi\tau_{\eta}(\xi)^2}{4\sigma}
\right)
\erfc\left(R\rme^{-\rmi\pi/4}\sqrt{-\sigma}+
\frac{\tau_\eta(\xi)\rme^{\rmi\pi/4}}{2\sqrt{-\sigma}}\right)\rmd\xi 
\end{equation*}
is uniform for $R>1$, by the Wierstrass $M$-test.

In the case $\sigma=0$ we have the simpler integral
\begin{align*}
  \int_1^{R^2} \exp(-\sqrt{\beta}\tau_\eta(\xi))\rmd\beta =& 
\frac{2}{\tau_\eta(\xi)^2}\left( \tau_\eta(\xi) \left( \rme^{-\tau_\eta(\xi)}
-R\rme^{-R\tau_\eta(\xi)}\right)\right. \\
 & +\left.\left( \rme^{-\tau_\eta(\xi)}-\rme^{-R\tau_\eta(\xi)}\right)\right).
\end{align*}
The integral with respect to $\xi$ then converges uniformly by the
Wierstrass $M$-test, since
\begin{equation*}
  |\exp(-R\tau_\eta(\xi))|\leq\exp(-\Re\;\tau_\eta(\xi))
\end{equation*}
and
\begin{equation*}
|R\exp(-R\tau_\eta(\xi))|\leq\frac{2}{\tau^{\ast}\rme}
\exp(-\textstyle\frac{1}{2}
\Re\;\tau_\eta(\xi))
\end{equation*}
for $R>1$.
\finire

The following theorem from \S11.55.III of \cite{S} describes 
criteria which permit the
interchange of order of two improper integrals.
\begin{theorem} \label{thm:app:1}
  Let $f(x,\alpha)$ be continuous in $\alpha_1\leq\alpha\leq\alpha_2$ and 
$c\leq x\leq d$, 
where both $\alpha_2$ and $d$ may be arbitrarily large, and;
\begin{equation*}
\begin{array}{rll}
\mbox{i)}& \displaystyle 
\int_c^{\infty} f(x,\alpha)\rmd x &\mbox{be uniformly convergent for 
$\alpha\in[\alpha_1,\alpha_2]$,}\\
  \mbox{ii)}& \displaystyle
\int_{\alpha_1}^{\infty} f(x,\alpha)\rmd\alpha&\mbox{be uniformly convergent
for $x\in[c,d]$,}\\
\mbox{iii)}& \displaystyle
\int_c^{\infty}\int_{\alpha_1}^{R} f(x,\alpha)\rmd\alpha \rmd x &
\mbox{be uniformly convergent for $R\in[\alpha_1,\infty]$.}
\end{array}
\end{equation*}
then
\begin{equation*}
 \int_{\alpha_1}^{\infty}\int_c^{\infty} f(x,\alpha)\rmd x\rmd\alpha =
\int_c^{\infty} \int_{\alpha_1}^{\infty} f(x,\alpha)\rmd \alpha\rmd x.
\end{equation*}
\end{theorem}

Applying theorem \ref{thm:app:1} to the integral in \eqref{eq:3:12} 
allows us to conclude the following.
\begin{proposition} \label{prop:a}
\begin{equation*}
  \int_0^{\infty} \int_0^{\infty} P_{\eta}(\xi)\exp(-\sqrt{\beta}\tau_\eta(\xi)
-\rmi\sigma\beta)\rmd\xi\rmd\beta =
\int_0^{\infty} \int_0^{\infty} P_{\eta}(\xi)\exp(-\sqrt{\beta}\tau_\eta(\xi)
-\rmi\sigma\beta) \rmd\beta\rmd\xi.
\end{equation*}
\end{proposition}
\dimostrazione
This follows from theorem \ref{thm:app:1} with lemmas \ref{lem:app:3}, 
\ref{lem:app:4} and \ref{lem:app:5}, together with corollary \ref{corr:app:1}.
\finire
\section{Appendix: Simplification of \eqref{eq:greg:greg}}
We here consider some technical points that arise in section \ref{sec:6}.
\begin{lemma} \label{lem:app:6}
  The integral 
  \begin{equation} \label{eq:tre}
   \Re \int_0^{\infty} P_\eta(\xi)\left(\frac{\rme^{3\rmi\pi/4}\tau_\eta(\xi)}
{2(-\sigma)^{3/2}}w\left(\frac{\rme^{3\rmi\pi/4}\tau_\eta(\xi)}{2\sqrt{-\sigma}}\right)\right)\rmd\xi
  \end{equation}
is uniformly convergent for $\sigma\in[-R^2,0]$ for any $R>0$.
\end{lemma}
\dimostrazione
Expanding the $w$ function, using lemma \ref{lem:w},
\begin{equation*}
\frac{\rme^{3\pi\rmi/4}\tau_\eta(\xi)}{2(-\sigma)^{3/2}}
  w\left(\frac{\rme^{3\rmi\pi/4}\tau_\eta(\xi)}{2\sqrt{-\sigma}}\right)
=\frac{1}{\rmi\sigma}+
\Ord\left(\frac{1}{\tau_\eta(\xi)^2}\right)
\end{equation*}
as $\xi\to\infty$ where the implied constant is independent of 
$\sigma\in[-R,0]$.
Since $\int_0^\infty P_\eta(\xi)\rmd\xi=\alpha$ the leading order term
in the expansion of \eqref{eq:tre} has zero real part, and the integral of the
remainder converges since $\tau_\eta(\xi)^{-2}\sim (\alpha\xi)^{-2}$ as 
$\xi\to\infty$.
\finire

\begin{proposition} \label{prop:app:2}
  We have
  \begin{equation*}
\lim_{R\to\infty}    \int_0^{\infty} \int_{-\infty}^{-R^2}P_\eta(\xi)
\left(\frac{\rme^{3\rmi\pi/4}\tau_\eta(\xi)}
{2(-\sigma)^{3/2}}w\left(\frac{\rme^{3\rmi\pi/4}\tau_\eta(\xi)}{2\sqrt{-\sigma}}\right)\right)\rmd\sigma\rmd\xi=0.
  \end{equation*}
\end{proposition}
\dimostrazione
We make the substitution $2p=(-\sigma)^{-1/2}$ to give
\begin{align*}
   \int_{-\infty}^{-R^2}
\frac{\rme^{3\rmi\pi/4}\tau_\eta(\xi)}
{2(-\sigma)^{3/2}}w\left(\frac{\rme^{3\rmi\pi/4}\tau_\eta(\xi)}{2\sqrt{-\sigma}}\right)\rmd\sigma&=\int_0^{1/2R} 2\rme^{3\rmi\pi/4}\tau_{\eta}(\xi)
w(\rme^{3\rmi\pi/4}\tau_{\eta}(\xi)p)\rmd p \\
&=\int_{\gamma_{\xi,R}} 2w(t)\rmd t
\end{align*}
where $t\in\C$ follows the contour $\gamma_{\xi,R}$ connecting 0 to
$\displaystyle \frac{\rme^{3\rmi\pi/4}\tau_{\eta}(\xi)}{2R}$. 
Since $w$ is an analytic function, we can write
\begin{equation*}
  \int_{-\infty}^{-R^2}
\frac{\rme^{3\rmi\pi/4}\tau_\eta(\xi)}
{2(-\sigma)^{3/2}}w\left(\frac{\rme^{3\rmi\pi/4}\tau_\eta(\xi)}
{2\sqrt{-\sigma}}\right)\rmd\sigma=2W\left(\frac{\rme^{3\rmi\pi/4}
\tau_{\eta}(\xi)}{2R} \right)
\end{equation*}
where $W$ is the antiderivative of $w$ satisfying
\begin{equation*}
  \frac{\rmd W}{\rmd z}=w(z)\qquad\mbox{and}\qquad W(0)=0.
\end{equation*}
By making the substitution $\xi\mapsto R\xi$, we see that
\begin{align}
  \int_0^\infty P_\eta({\xi})W\left(\frac{\rme^{3\rmi\pi/4}
\tau_\eta({\xi})}{2R} \right){\rmd\nu}
=&\frac{R}{2}\int_0^1 P_\eta(R\sqrt{\nu})W\left(\frac{\rme^{3\rmi\pi/4}
\tau_{\eta}(R\sqrt{\nu})}{2R} \right)\frac{\rmd\nu}{\sqrt{\nu}} \nonumber \\
&+R\int_1^\infty P_\eta(R\xi)W\left(\frac{\rme^{3\rmi\pi/4}
\tau_{\eta}(R\xi)}{2R} \right)\rmd\xi \label{eq:quattro}
\end{align}
where we have, additionally, split the range of integration into two regimes
and the first integral made the substitution $\nu=\xi^2$.
For the first integral in \eqref{eq:quattro} we consider
\begin{equation*}
  \int_0^1 \frac{R}2\exp\left(\frac{\rmi\pi}{4}-\frac{\rmi R^2\nu}{4}\right)
W\left(\frac{\rme^{3\rmi\pi/4}\tau_{\eta}(R\sqrt{\nu})}{2R} \right)
\frac{\rmd\nu}{\sqrt{\nu}}
\end{equation*}
which comes from the first term of $P_\eta(R\sqrt{\nu})$.
The second term of $P_\eta$ can be handled in the same way.
Differentiating,
\begin{equation}  \label{eq:cinque}
  \frac{\rmd}{\rmd\nu}\left[W\left(\frac{\rme^{3\rmi\pi/4}
\tau_{\eta}(R\sqrt{\nu})}{2R} \right) \right]
=\frac{\rme^{3\rmi\pi/4}}{4\sqrt{\nu}}
w\left(\frac{\rme^{3\rmi\pi/4}\tau_{\eta}(R\sqrt{\nu})}{2R} \right)
\tau_{\eta}^{\prime}(R\sqrt{\nu}).
\end{equation}
Since $\displaystyle \frac{\rmd\tau_\eta}{\rmd\xi}$ is bounded for $\xi\in\R$,
we deduce from \eqref{eq:cinque} that there exists a constant $K$ independent
of $R$ such that
\begin{equation}  \label{eq:sei}
  \left|\frac{\rmd}{\rmd\nu}\left[W\left(\frac{\rme^{3\rmi\pi/4}
\tau_{\eta}(R\sqrt{\nu})}{2R} \right) \right]\right| \leq
\frac{K}{\sqrt{\nu}}.
\end{equation}
Let
\begin{equation*}
\psi(\nu)\coloneq{-\sqrt{\pi}}\erfc\left(\frac{R\rme^{\rmi\pi/4}
\sqrt{\nu}}{2} \right)  
\end{equation*}
which satisfies
\begin{equation*}
  \frac{\rmd\psi}{\rmd\nu}=
\frac{R}{2\sqrt{\nu}}\exp\left(\frac{\rmi\pi}{4}-\frac{\rmi R^2\nu}{4}\right).
\end{equation*}
We can then use integration by parts,
\begin{align}
  \int_0^1 \frac{R}2&\exp\left(\frac{\rmi\pi}{4}-\frac{\rmi R^2\nu}{4}\right)
W\left(\frac{\rme^{3\rmi\pi/4}\tau_{\eta}(R\sqrt{\nu})}{2R} \right)
\frac{\rmd\nu}{\sqrt{\nu}} \\
&=\left[\psi(\nu)W\left(\frac{\rme^{3\rmi\pi/4}
\tau_{\eta}(R\sqrt{\nu})}{2R} \right) \right]_0^1 
-\int_0^1 \psi(\nu)\frac{\rmd}{\rmd\nu}\left[W\left(\frac{\rme^{3\rmi\pi/4}
\tau_{\eta}(R\sqrt{\nu})}{2R} \right) \right]\rmd\nu \nonumber \\
&\to 0 \label{eq:sette}
\end{align}
as $R\to\infty$, since
\begin{equation*}
  W\left(\frac{\rme^{3\rmi\pi/4}
\tau_{\eta}(0)}{2R} \right)\to 0
\end{equation*}
and
\begin{equation*}
  \erfc\left(\frac{R\rme^{\rmi\pi/4}}{2} \right)\to0
\end{equation*}
and the fact that the final integral in \eqref{eq:sette} converges uniformly
by \eqref{eq:sei}.

For the second integral in \eqref{eq:quattro} we apply Taylor's
theorem and lemma \ref{lem:app:1} to get
\begin{equation*}
  W\left(\frac{\rme^{3\rmi\pi/4}\tau_{\eta}(R\xi)}{2R} \right)
=W\left(\frac{\rme^{3\rmi\pi/4}\alpha\xi}{2} \right)+
\Ord\left(\frac{1}{R^3\xi^2}
\right).
\end{equation*}
This gives
\begin{equation*}
  R\int_1^\infty P_\eta(R\xi)W\left(\frac{\rme^{3\rmi\pi/4}
\tau_{\eta}(R\xi)}{2R} \right)\rmd\xi
=R\int_1^\infty P_\eta(R\xi)W\left(\frac{\rme^{3\rmi\pi/4}
\alpha\xi}{2} \right)\rmd\xi+\Ord(R^{-2})
\end{equation*}
as $R\to\infty$. The integral which remains is of a form for which
the asymptotic series may be derived by the method of repeated 
integration-by-parts \cite{BH} to see that this contribution also
vanishes in the limit $R\to\infty$. \finire


\begin{thebibliography}{9999999}
\bibitem[AS]{AS}{Abramowitz M and Stegun I A 1965 {\it Handbook of 
mathematical functions} (Dover)}
\bibitem[BSS]{BSS}{B\"acker A, Schubert R and Sifter P 1998 Rate of quantum
ergodicity in Euclidean billiards {\it Phys.\ Rev.\ E} {\bf 57} 5425--5447,
erratum {\it ibid.} {\bf 58} 5192} 
\bibitem[BG]{BG}{Barra F and Gaspard P 2000 {On the level spacing 
distribution in quantum graphs} {\it J.\ Stat.\ Phys.\ }{\bf 101} 283--319}
\bibitem[B]{Berk}{Berkolaiko G Form factor for large quantum graphs: 
evaluating orbits with time-reversal {\em preprint} available at
{\tt arXiv:nlin.CD/0305009}}
\bibitem[BBK]{BBK}{Berkolaiko G, Bogomolny E B and Keating J P 2001 Star
graphs and \v{S}eba billiards {\it J.\ Phys.\ A} {\bf 34} 335--350}
\bibitem[BK]{BK}{Berkolaiko G and Keating J P 1999 Two-point spectral
correlations for star graphs {\it J.\ Phys.\ A} {\bf 32} 7827--7841}
\bibitem[BSW1]{BSW1}{Berkolaiko G, Schanz H and Whitney R S 2002 {Leading 
off-diagonal correction to the form factor of large graphs} {\it Phys.\
Rev.\ Lett.} {\bf 82} art.\ no.\ 104101}
\bibitem[BSW2]{BSW2}{Berkolaiko G, Schanz H and Whitney R S 2003 Form factor 
for a family of quantum graphs: An expansion to third order 
{\em J.\ Phys.\ A} {\bf 36} 8373--8392}
\bibitem[BKW]{BKW}{Berkolaiko G, Keating J P and Winn B Intermediate 
wave-function statistics {\em preprint} available at
{\tt arXiv:nlin.CD/0304034}}
\bibitem[Be1]{B}{Berry M V 1977 {Regular and irregular semiclassical 
wavefunctions} {\it J.\ Phys.\ A} {\bf 10} 2083--2091}
\bibitem[Be2]{B2}{Berry M V 1989 Quantum scars of classical closed orbits in 
phase space {\it Proc.\ Roy.\ Soc.\ Lond.\ A} {\bf 423} 219--231}
\bibitem[BlHa]{BH}{Bleistein N and Handlesman R A 1986 {\it Asymptotic
expansions of integrals} (Dover)}
\bibitem[Bo]{Bog}{Bogomolny E B 1988 Smoothed wave functions of chaotic
quantum systems {\it Physica D} {\bf 31} 169--189}
\bibitem[BH1]{BoHa1}{Bolte J and Harrison J 2003 Spectral statistics for
the Dirac operator on graphs {\it J.\ Phys.\ A} {\bf 36} 2747--2769}
\bibitem[BH2]{BoHa2}{Bolte J and Harrison J 2003 The spin contribution to
the form factor of quantum graphs {\em J.\ Phys.\ A} {\bf 36} L433--L440}
\bibitem[CdV]{CdV}{Colin de~Verdi\`ere Y 1985 {Ergodicit\'e et fonctions
propres du Laplacien} {\it Commun.\ Math.\ Phys} {\bf 102} 497--502}
\bibitem[CDM]{CDM}{Comtet A, Desbois J and Majumdar S N 2002 The local time 
distribution of a  particle diffusing on a graph {\it J.\ Phys.\ A} {\bf 35}
687--694}
\bibitem[D]{D}{Desbois J 2002 Occupation times distribution for Brownian
motion on graphs {\it J.\ Phys.\ A} {\bf 35} 673--678}
\bibitem[DS]{DS}{Dunford N and Schwartz J T 1963 {\em Linear Operators Part II:
Spectral Theory} (Interscience Publishers)}
\bibitem[FNdB]{FNdB}{Faure F, Nonnenmacher S and de~Bi\`evre S
Scarred eigenstates for quantum cat maps of minimal periods {\em preprint}
available at {\tt arXiv:nlin.CD/0207060}}
\bibitem[GL]{GL}{G\'erard P and Leichtnam E 1993 {Ergodic properties of the
eigenfunctions for the Dirichlet problem} {\it Duke Math. J.} {\bf 71}
559--607}
\bibitem[GR]{GR}{Gradshteyn I S and Ryzhik I M 1965 {\it Tables of integrals,
series, and products} (Academic Press)}
\bibitem[GSW]{GSW}{Gnutzmann S, Smilansky U and Weber J Nodal domains on 
quantum graphs {\em preprint} available at {\tt arXiv:nlin.CD/0305020}}
\bibitem[H]{H}{Heller E J 1984 {Bound-state eigenfunctions of classically 
chaotic Hamiltonian systems: scars of periodic orbits} 
{\it Phys.\ Rev.\ Lett.} {\bf 53} 1515--1518}
\bibitem[K1]{Ka}{Kaplan L 1999 Scars in quantum chaotic wavefunctions
{\it Nonlinearity} {\bf 12} R1--R40}
\bibitem[K2]{K2}{Kaplan L 2001 Eigenstate structure in graphs and
disordered lattices {\it Phys.\ Rev.\ E} {\bf 64} art.\ no. 036225}
\bibitem[KH]{KH}{Kaplan L and Heller E J 1998 Linear and nonlinear theory
of eigenfunction scars {\it Ann. Phys.} {\bf 264} 171--206}
\bibitem[Ke]{K}{Keating J P 1991 The cat maps: quantum mechanics and 
classical motion {\it Nonlinearity} {\bf 4} 309--341}
\bibitem[KMW]{KMW}{Keating J P, Marklof J and Winn B  {Value distribution
of the eigenfunctions and spectral determinants of quantum star graphs.}
{To appear in {\it Commun.\ Math.\ Phys.}} Preprint available at
{\tt arXiv:math-ph/0210060}}
\bibitem[KP]{KP}{Keating J P and Prado S D 2001 Orbit bifurcations and the
scarring of wavefunctions {\it Proc.\ Roy.\ Soc.\ Lond.\ A} {\bf 457}
1855--1872}
\bibitem[KS1]{KS1}{Kottos T and Smilansky U 1997 Quantum Chaos on graphs
{\it Phys.\ Rev.\ Lett.} {\bf 79} 4794--4797}
\bibitem[KS2]{KS2}{Kottos T and Smilansky U 1999 Periodic orbit theory and
spectral statistics for quantum graphs {\it Ann.\ Phys.} {\bf 274} 76--124}
\bibitem[KS3]{KS3}{Kottos T and Smilansky U 2000 Chaotic scattering on graphs
{\it Phys.\ Rev.\ Lett.} {\bf 85} 968--971}
\bibitem[KS4]{KS4}{Kottos T and Smilansky U 2003 Quantum graphs: a simple
model for chaotic scattering {\it J.\ Phys.\ A} {\bf 36} 3501--3524}
\bibitem[PTZ]{PTZ}{Pako\'nski P, Tanner G and \.{Z}yczkowski K 2003 Families
of line-graphs and their quantization {\it J.\ Stat.\ Phys.} {\bf 111}
1331--1351}
\bibitem[S]{Sc}{Schnirelmann A 1974 {Ergodic properties of eigenfuncions} 
{\it Usp.\ Math.\ Nauk.} {\bf 29} 181--182}
\bibitem[St]{S}{Stewart C A 1940 {\it Advanced Calculus} (Methuen)}
\bibitem[Se]{Se}{\v{S}eba P 1990 Wave chaos in singular quantum billiards
{\it Phys.\ Rev.\ Lett.} {\bf 64} 1855--1858}
\bibitem[SK]{SK}{Schanz H and Kottos T 2003 {Scars on quantum networks
ignore the Lyapunov exponent} {\em Phys.\ Rev.\ Lett.} {\bf 90}
art.\ no.\ 234101}
\bibitem[T]{T}{Tanner G 2001 Unitary stochastic matrix ensembles and
spectral statistics {\it J.\ Phys.\ A} {\bf 34} 369--383}
\bibitem[TM]{TM}{Texier C and Montambaux G 2001 Scattering theory on graphs
{\it J.\ Phys.\ A} {\bf 34} 10307--10326}
\bibitem[V]{V}{Voros A 1979 {Semi-classical ergodicity of quantum eigenstates
in the Wigner representation}, in {\it Stochastic behaviour in classical
and quantum Hamiltonian systems} (Springer-Verlag) pp.\ 326--333}
\bibitem[W]{W}{Weyl H 1916 \"{U}ber die Gleichverteilung von Zahlen mod. Eins
{\it Math.\ Ann.\ }{\bf 77} 313--352}
\bibitem[Z]{Z}{Zelditch S 1987 {Uniform distribution of the eigenfunctions
on compact hyperbolic surfaces} {\it Duke Math.\ J.} {\bf 55} 919--941}
\end{thebibliography}
\end{document}